\documentclass[aps,notitlepage,superscriptaddress,nofootinbib,twocolumn,floatfix,10pt,pra]{revtex4}

\usepackage{graphicx}
\usepackage{amsmath,amssymb,amsfonts,amsthm,stmaryrd,mathtools,bm,physics}
\usepackage{color}
\usepackage{tikz}
\usepackage[normalem]{ulem}
\usepackage{braket}
\usepackage{fix-cm}
\allowdisplaybreaks[1]

\def\be#1\ee{\begin{align}#1\end{align}}

\def\be{\begin{eqnarray}}
\def\ee{\end{eqnarray}}

\def\rmd{\mathrm{d}}

\usepackage[bookmarks,linktocpage, colorlinks=true, plainpages = false, citecolor = treegreen,  linkcolor=darkblue, urlcolor = darkblue, filecolor = blue]{hyperref} 
\usepackage{comment}
\usepackage{xargs}
\usepackage{todonotes}

\newcommandx{\hgl}[2][1=]{\todo[linecolor=blue,backgroundcolor=blue!25,bordercolor=blue,#1]{#2}}

\definecolor{darkblue}{rgb}{0., 0.4, 0.8}
\definecolor{treegreen}{rgb}{0., 0.7, 0.3}

\def\be#1\ee{\begin{equation}\begin{aligned}#1\end{aligned}\end{equation}}

\begin{document}

\title{Quantum induced shock dynamics in gravitational collapse: insights from effective models and numerical frameworks}

\author{Hongguang Liu}
\email{liuhongguang@westlake.edu.cn} 
\affiliation{Institute for Theoretical Sciences, Westlake University, Hangzhou 310030, China}
\affiliation{Institute of Natural Sciences, Westlake Institute for Advanced Study, Hangzhou 310024, China}
\author{Dongxue Qu}
\email{dqu@perimeterinstitute.ca} 
\affiliation{Perimeter Institute for Theoretical Physics, 31 Caroline St N, N2L 2Y5 Waterloo, ON, Canada}

\begin{abstract}
We explore the formation and evolution of shock waves in spherically symmetric gravitational collapse within a Loop Quantum Gravity (LQG) inspired effective framework. In this setting, the classical singularities are replaced by quantum-induced shell-crossing singularities, which are resolved through weak solutions such as shock waves. By formulating the dynamics in a generalized Painlevé–Gullstrand coordinate system, we derive a first-order partial differential equation that governs the propagation of the shock surface, while enforcing metric continuity via thin-shell junction conditions. To handle the non-trivial square-root structures and source terms that arise in these equations, we develop a novel numerical scheme capable of simulating quantum-corrected spacetime dynamics. Our results show that for small mass black holes near the Planck scale, the shock surface remains timelike and is shielded behind both inner and outer horizons. In the long-time limit, the shock accumulates the entire mass of the collapsing star. In contrast, for larger black hole masses, the shock surface develops spacelike segments, indicating a transition in the effective dynamics driven by quantum effects. The framework also reveals discontinuities in curvature invariants across the shock surface, which can be traced back to stress-energy redistributions caused by quantum effects. Overall, the proposed computational framework provides a general tool for modeling quantum-corrected gravitational collapse and offers new insights into black hole formations, singularity resolution, and the interplay between quantum geometry effects and effective spacetime structures.
\end{abstract}
\maketitle 

\section{Introduction}
Black holes are expected to evolve into spacetime regimes where strong quantum gravity effects dominate, requiring a departure from classical general relativity (GR) to describe their late-stage evolution. Within the framework of Loop Quantum Gravity (LQG), many quantum black hole models have been developed \cite{Ashtekar:2005qt,Modesto:2005zm,Boehmer:2007ket,Chiou:2012pg,Gambini:2013hna,Corichi:2015xia,Dadhich:2015ora,Olmedo:2017lvt,Ashtekar:2018lag,BenAchour:2018khr,Han:2020uhb,Ashtekar:2023cod, Bojowald:2008ja, Tibrewala:2012xb, Kelly:2020lec, Husain:2021ojz, Husain:2022gwp,Han:2022rsx, Giesel:2023tsj, Giesel:2023hys, Fazzini:2023ova, Giesel:2024mps}. These models provide a theoretical framework for studying gravitational collapse and regular black holes by incorporating quantum geometric effects, such as discrete spacetime structures and polymerized modifications to the Hamiltonian dynamics. A key feature emerging in LQG-inspired scenarios is the appearance of a quantum bounce near the classical singularity. In various bouncing models, LQG effects can transform a classical central shell-focusing singularity into a quantum-driven shell-crossing singularity \cite{Fazzini:2023ova, Giesel:2024mps}. The latter is a weak singularity that can be resolved by introducing shock wave solutions. These shock waves are natural weak solutions to the dynamical equations of effective models, which take the form of a system of first-order quasi-linear equations \cite{Giesel:2023hys,Fazzini:2023ova}.

It is well known that for systems of first-order quasi-linear equations, weak solutions such as shock waves are not unique. Their form depends on the choice of variables in the equations. A change of variable can change the behavior of the weak solution \cite{evans10,Fazzini:2025hsf}. Such phenomenon also happens in classical GR, for example in dust models with inhomogeneous profiles \cite{Nolan:2003wp}. As a result, a key challenge in deriving dust-driven dynamics, especially in effective theories for regular black holes, is to determine which shock solution is physically meaningful. Addressing this problem requires more than just the dynamical equations. Additional input from physical principles or matter fields is needed. In this paper, we explore this issue in detail.

A natural way to constrain potential shock structures is to impose a weak continuity condition on the metric field. In classical GR, this idea appears in the form of the thin shell formalism \cite{Mars:1993mj,Senovilla:2018hrw}, which serves as a weak solution framework for Einstein's equations that allows for a generalized formulation in a distributional sense. In particular, the weak continuity of the metric field ensures the continuity of the first fundamental form across the junction (shock) surface. When this condition is expressed in the Painlevé–Gullstrand (PG) coordinate system, it leads to an ordinary differential equation (ODE). The solution, $r_s(t)$, describes the trajectory of the junction surface once an initial point is specified. Consequently, these jump conditions impose constraints that must be satisfied by the weak solutions of the first-order equations and help determine the correct form of those equations. Note that in this formulism we allow the second fundamental forms to be discontinuous such that a singular distributional matter distribution (thin shell of matter) are allowed on the shock surface. This is exact what happens in classical General Relativity as shown in \cite{Tegai:2011qf} and very recently in \cite{Husain:2025wrh} \footnote{The work in \cite{Husain:2025wrh} which appeared as this work was being completed, also investigates the uniqueness of shock solutions. However, it exclusively focuses on classical general relativity, whereas our framework incorporates all effective models defined in \cite{Giesel:2024mps}, including both classical and LQG-inspired cases.}.

In this paper, we illustrate the method explicitly in the marginally bound case, using an LQG-inspired symmetric bounce model as a concrete example. We identify a unique set of variables for the first-order PDEs such that they are compatible with the thin-shell formalism, meaning that their jump conditions agree with the continuity of the first fundamental form. This approach is broadly applicable to all LTB models of dust collapse discussed in \cite{Giesel:2024mps}, including those in classical general relativity, Thiemann-regularized asymmetric bouncing models, and geometries with a regular center, such as the Hayward or Bardeen metrics. The symmetric bounce model inspired by LQG admits analytical solutions along characteristic lines, as the jump condition can be directly integrated using LTB solutions \cite{Giesel:2023hys}. In more complex cases, however, a numerical approach becomes necessary. In this work, we develop a numerical scheme that directly integrates these first-order equations.

Compared with the equations provided in \cite{Kelly:2020lec}, our new PDE system compatible with the thin shell formalism is reformulated using $N^x$ as the primary variable instead of the connection $b$. As shown in \cite{Giesel:2023hys,Giesel:2024mps}, defining variables such as $N^x$ requires introducing a lift that selects appropriate branches in bouncing models due to their multivalued behavior in phase space. In symmetric bouncing models, for instance, this lift corresponds to choosing a branch of the square-root function. For our numerical implementation, we developed a method to automatically account for this feature.

Our approach shows that choosing variables compatible with the thin-shell formalism enforces the continuity of the metric across the shock surface, unlike the formulation in \cite{Kelly:2020lec}, which allows discontinuous metrics. This framework enables a systematic analysis of the shock surface's dynamics, including its causal structure, and provides greater insights into gravitational collapse.

In the context of an LQG-inspired black hole model with symmetric bounce, our analysis shows that for a quantum black hole with mass close to its minimal allowable value (given by \( M_c = \frac{8\zeta}{3\sqrt{3}G} \), where \(\zeta\) is a quantization parameter and \( G \) is the gravitational constant, corresponding to Planckian scales), the shock surface remains timelike. This identifies it as a physically meaningful shock wave. Crucially, this shock lies entirely within the untrapped region interior to the inner horizon of the black hole, and is therefore not observable by an  external observer.

Moreover, the long-term evolution of the shock wave asymptotically approaches the minimal radius prescribed by the model's bounce condition. In this late-time regime, all of the collapsing star’s mass becomes concentrated within the shock, which stabilizes at the minimal radius, whose interior becomes a flat spacetime region.

However, when the mass increases beyond a critical threshold, the shock surface develops spacelike segments that intersect the black hole’s inner horizon. This results give rise to non-physical regions within the shock wave. Although the long-term behavior of the shock is similar to that in smaller black holes, the appearance of these unphysical regions highlights fundamental limitations of current effective descriptions, indicating that quantum tunneling effects may be essential at such mass scales. These findings also emphasize the critical role of black hole evaporation in understanding spacetime geometry of large black hole: only after a black holes has evaporated to near-Planckian masses does the effective dynamics predict a physically viable shock wave.


The paper is structured as follows: In Section \ref{LQG Black hole model}, we introduce the LQG black hole model and its effective dynamics, highlighting shell-crossing singularities and their resolution using jump conditions that compatiable with thin shell junction conditions derived from jump conditions. Section \ref{Numerical results} outlines the numerical methods used to solve the PDEs, including discretization techniques and shock-capturing schemes. Section \ref{Implementation} details the numerical implementation and presents results on shock wave formation. Based on these results, we investigate the causal signature of the shock surface, examine the formation of trapped and anti-trapped regions, and analyze the behavior of curvature scalars on both sides of the shock surface. Finally, Section \ref{conclusion} summarizes our findings and discusses their implications for quantum gravity and black hole physics.

\section{LQG Black hole model}\label{LQG Black hole model}
We start with the general metric for the spherically symmetric spacetime:
\be 
\rmd s^2 = -\rmd t^2 + \frac{(E^{\phi})^2}{|E^x|}\left(\rmd x+N^x\rmd t \right)^2 +|E^x|\rmd \Omega^2
\ee 
where $\rmd \Omega^2=\rmd \theta^2+\sin^2\theta\rmd \phi^2$. The functions $E^{\phi}(t,x)$ and $E^x(t,x)$ are related to the densitized triad field on the spatial slice, inspired by LQG \cite{Ashtekar:2005qt,Modesto:2005zm,Boehmer:2007ket,Chiou:2012pg,Gambini:2013hna,Corichi:2015xia,Dadhich:2015ora,Olmedo:2017lvt,Ashtekar:2018lag,Han:2020uhb,Ashtekar:2023cod}. In the case of a dust-filled universe, the dynamics are determined by a physical Hamiltonian, with the time direction gauge-fixed using a dust field. We focus on the effective dynamics of an LQG-inspired model, where the Hamiltonian is given by \cite{Giesel:2023tsj,Giesel:2023hys}
\be\label{eq:eff_Ham_m}
\mathcal{H}^{\Delta}_{pri} =& \int \dd{x} \,\qty( {C}^{\Delta} + N^x C_x), \quad
C_x = \frac{1}{G} \left( E^{\phi} K_{\phi}' - K_x E^x{}' \right),\\
C^{\Delta} =&- \frac{E^{\phi} \sqrt{E^x}}{2G} \bigg[\frac{3}{\zeta^2} \sin^2\left(\frac{\zeta K_{\phi}}{\sqrt{{E^x}}}\right) \\
&+\frac{ (2 {E^x} K_x-{E^{\phi}} K_{\phi})}{\zeta \sqrt{{E^x}} {E^{\phi}}}  \sin \left(\frac{2 \zeta K_{\phi}}{\sqrt{{E^x}}}\right)+\\
&\quad\quad+ \frac{1 -\qty(\frac{  {{E^x}}'}{2{{E^{\phi}}} })^2}{E^x}  - \frac{2}{E^\phi}\Big(\frac{  {E^x}'}{2{{E^{\phi}}} }\Big)'\bigg],
\ee
Here, $K_{\phi}$ and $K_x$ are connection fields that form canonical pairs with $E^{\phi}$ and $E^x$, respectively, such that $\{ K_{\phi}, E^{\phi} \}= \{ K_{x}, E^{x} \} = G$. The above dynamics satisfy the Lemaître-Tolman-Bondi (LTB) conditions  \cite{Bojowald:2008ja,Bojowald:2009ih, Giesel:2023tsj}
\begin{align}
    E^{\phi} = \frac{{E^x}'}{2\sqrt{1+\epsilon(x)}}, \quad  K_x = \frac{{K_{\phi}}'}{2\sqrt{1+\epsilon(x)}}
\end{align}
where $\epsilon(x)$ is time-independent. It tells us if the spacetime is bound ($\epsilon < 0 $), marginally bound ($\epsilon = 0 $) , or unbound ($\epsilon > 0 $). This condition encodes the system into LTB coordinates. The metric becomes:
\begin{align}
    \mathrm{d}s^2 = - \mathrm{d}t^2 + \frac{(R')^2}{1+\epsilon(x)} \mathrm{d}x^2 + R^2 \mathrm{d} \Omega^2,
\end{align}
where $R(t,x) \equiv \sqrt{E^x}$ is the areal radius. In this coordinates, the dynamics become simpler and follow cosmological dynamics: 
\begin{align}\label{eq:ltbequations}
  \frac{\dot{R}}{R} =&\frac{\sin(2 \zeta b)}{ 2 \zeta}, \quad \dot{K_{\phi}} = -\frac{\frac{3 \sin\left( \zeta b \right)^2}{\zeta^2 } - 2 b \frac{\sin(2 \zeta b)}{ 2 \zeta} - \epsilon}{2 R } \ ,
\end{align}
where $b \equiv \frac{K_{\phi}}{R}$ is the $\bar{\mu}$-scheme version of the connection $K_{\phi}$. The quantum parameter $\zeta$ is proportional to the Planck length $\sqrt{\hbar}$. These equations can be written as a modified Friedmann equation for each $x$, like in loop quantum cosmology (LQC):
\begin{eqnarray}
\label{Friedmann from homogeneous reduction}
\frac{\dot{R}^2}{R^2}(x) =\left( \frac{\kappa  \rho }{6} + \frac{\epsilon}{R^2} \right) \left( 1 - \zeta^2\left( \frac{\kappa  \rho }{6} + \frac{\epsilon}{R^2} \right) \right)(x) \,,
\end{eqnarray}
with $\rho \equiv \frac{3 }{4 \pi R^3} M'(x) $, where $M(x)$ is a conserved quantity:
\begin{align}
    2 G M(x) \equiv \partial_x C^{\Delta} = R \left( R^2 \frac{\sin(\zeta b)^2}{\zeta^2} + \epsilon \right).
\end{align}
Here, $M(x)$ is known as the mass function which represents the total dust energy inside the shell at radius $x$. In classical general relativity (GR), this mass function coincides with the Misner-Sharp mass \cite{Cahill1970SphericalSA},  which measures the total energy in the spherically symmetric spacetime. 

In the marginally bound case ($\epsilon=0$), the general solution to (\ref{Friedmann from homogeneous reduction}) is: 
\begin{eqnarray}
    R(t,x) =  \left(2 G M(x) \right)^\frac{1}{3}\left( \zeta^2 + \frac{9}{4} z^2 \right)^\frac{1}{3}, \quad z = s(x) - t\,.
\end{eqnarray}
where $s(x)$ is a free integration function. In the polymerized vacuum, where $M(x) = \text{const}$, we can set $s(x)=x$. For the homogeneous dust case, we choose $s(x)=0$.

Some studies show that the polymerized vacuum spacetime does not have \textit{shell-crossing singularity}. But when the dust density is non-uniform, such singularities will arise \cite{Fazzini:2023ova, Giesel:2024mps}. These singularities lead to a quadratic divergence of the Kretschmann scalar. For example, in the marginally bound case - where the dust moves along parabolic paths with no extra forces - the Kretschmann scalar is given by \cite{Giesel:2024mps}:
\begin{align}
    \mathcal{K}=\frac{\mathcal{A} }{\left(9 (s(x)-t)^2+4 \zeta^2 \right)^4 \mathcal{S}^2},    
\end{align}
where $\mathcal{A}$ depends on $M(x)$ and its derivative $M'(x)$, which describe the mass distribution of the dust. The denominator includes the term,
\be
\mathcal{S} = M'(x) \left[9 (s(x)-t)^2+4 \zeta^2 \right]+18 M(x) s'(x) \left[s(x)-t\right]\nonumber
\ee
which determines where the shell-crossing singularity appears. A singularity forms when $\mathcal{S} = 0$. Since this is a quadratic equation, if $9 M(x)^2 s'(x)^2-4 \zeta^2 M'(x)^2 \geq 0$, then for any given $x$, there is always a time $t$ when $\mathcal{S} = 0$. So the singularity is unavoidable in regions with non-uniform dust distribution. This happens, for example, in the region near the polymerized vacuum with very small fluctuations, where $M'(x)$ is sufficiently small and $M$ is approximately constant. But earlier work \cite{newman1986strengths,Joshi:2000fk} has shown that these singularities can be resolved by using \textit{weak solutions}—ones that may not be continuous but still satisfy the integral form of the partial differential equations (PDEs) of the system. 

\subsection{Junction condition}
To investigate the extension of the metric across singularities, a straightforward approach is to use junction conditions. This means identifying two surfaces, $\Sigma_{\pm}$, in regions before and after the singularity, and ensuring the continuity of the first fundamental form (the induced metric). In spherically symmetric spacetime, gluing the angular part of the metric always leads to the condition $R|_{\Sigma_+} = R|_{\Sigma_-}$ in LTB coordinates. Thus, using $r = R$ as a coordinate ensures coordinate continuity across the junction surface. It is possible that the junction surface is actually a shock if there is a non-trivial energy-momentum density on it. This shows that $(t,r)$ is a more suitable coordinate choice for studying the effect of weak solutions, which correspond to the Painlevé-Gullstrand (PG) coordinates:
\be 
\rmd s^2 = -\rmd t^2 + \frac{1}{1+\epsilon(t,r)}\left(\rmd r+N^x\rmd t \right)^2 +r^2\rmd \Omega^2.
\ee 
In this coordinate system, we have 
\begin{align}\label{eq:nx_in_m}
    N^x =& - \partial_t R(t,x) \\
    =&- \text{sgn}_b \sqrt{\frac{2 G M+\epsilon r}{r}\left( 1 - \frac{\zeta^2}{r^2} \frac{2 G M+\epsilon r}{r} \right) } \, . \nonumber
\end{align}
where the sign function $\text{sgn}_b \equiv \text{sign}( \partial_t R(t,x))$ defines a lift of the square root function to its covering space, as the square root of $\dot{R}$ in \eqref{Friedmann from homogeneous reduction}. This implies that whenever we cross a root of the square root function--at a bounce or recollapse--we must change the sign. Such lift is automatically captured if we use the variable variable $b$, since
\begin{align}\label{solNx_x^2}
    N^x = - \frac{r \sin \left( {2 \zeta b}\right)}{2 \zeta}\,.
\end{align}

For a hypersurface $\Sigma$ defined by $r=r_s(t)$, the induced metric becomes:
\begin{align} \label{eq:induced_metric}
\mathrm{d}S_{\Sigma}^2 &= \left( -1 + \frac{(N^x + r_s'(t))^2}{1+\epsilon} \right) \mathrm{d}t^2 + r_s^2 d\Omega^2 . 
\end{align} 
Continuity across the junction surface requires:
\begin{align}
    [r_s] = 0, \quad \Big[ \frac{1}{1+\epsilon} \left( N^x  + r_s'(t) \right)^2 \Big] =0 , 
\end{align}
where $[f]\equiv f_{+} - f_{-}$ denotes the jump across the surface. This leads to 
\begin{align}\label{eq:dynamical_junction}
    r_s'(t) =  - \frac{\sqrt{1+\epsilon_{+}} N^x_{-} + \sqrt{1+\epsilon_{-}} N^x_{+}}{ 2 \overline{\sqrt{1+\epsilon}} } ,
\end{align}
which is a first-order ordinary differential equation (ODE), where $\overline{f}\equiv \frac{f_{+} + f_{-}}{2}$. Given the initial condition $r(0)$ and the functions $N^x_{\pm}$ and $\epsilon_{\pm}$, \eqref{eq:dynamical_junction} uniquely determines the hypersurface $r = r_s(t)$. We notice that \eqref{eq:dynamical_junction} is independent of the detailed dynamics. Any model that admits an LTB solution—for example, the models in \cite{Giesel:2023tsj,Giesel:2024mps}—can use this approach.

It has been shown in \cite{Giesel:2023hys,Fazzini:2023ova, Giesel:2024mps} that the conservation laws for $M(x)$ and $\epsilon(x)$ in LTB coordinates generate the dynamical equations for $M(t,r)\equiv M(x(t,r))$ and $\epsilon(t,r) \equiv \epsilon(x(t,r))$ in PG coordinates. These are:
\begin{align}\label{eq:naive_set}
    \partial_t M  - N^x \partial_r M = 0,\\
    \partial_t \epsilon  - N^x \partial_r \epsilon = 0.
\end{align}
These equations are first-order quasi-linear PDEs that work for all LTB models studied in \cite{Giesel:2024mps}, from classical GR to quantum gravity-inspired models. The LTB equations in \eqref{eq:ltbequations} are exactly the characteristic equations (refer to Sec. \ref{Sec:Method of characteristics}) that describe the propagation of this system along its characteristic curves. Here, $N^x$ is treated as a function of $(M,\epsilon,r)$, and its exact form depends on the model. For instance, in LQG-inspired models with the Hamiltonian \eqref{eq:eff_Ham_m}, $N^x$ takes the form given by \eqref{eq:nx_in_m}. 

\subsection{Jump condition}
In the polymerized vacuum case, where $M$ is constant, the equation for $M$ becomes trivial and does not give any dynamics. To recover a meaningful evolution equation, we replace $M$ with $N^x$ in \eqref{eq:naive_set}, and rewrite the equation as:
\begin{align}\label{eq:new_eom_Nx}
    \partial_t N^x - \frac{1}{2}\partial_r (N^x)^2 = J(N^x, r) - \frac{\text{sgn}\sqrt{r^{2}-4 \zeta^2 (N^x)^2} \epsilon  }{2 r^2}. 
\end{align}
with 
\begin{align}
    J(N^x, r) = \frac{2 (N^x)^2}{r}  &+ \frac{3 \left(\text{sgn}\sqrt{r^{2}-4 \zeta^2 (N^x)^2}-r\right)}{4 \zeta^2}. 
\end{align}
Here, $\text{sgn}=\pm 1$ defines a lift of the square root, similar to $\text{sgn}_b$, which appears during obtaining $M$ from the the inversion of equation~\eqref{eq:nx_in_m}. 
The sign flips when the argument of the square root crosses zero.

It is well known that a first-order quasi-linear PDE in the form of a balance equation
\begin{align}\label{eq:1st_eq}
    \partial_t u + \partial_r A(u,r) = J(u,r)
\end{align}
can develop weak solutions that appear as shock waves, with their location determined by the jump condition \cite{evans10}:
\begin{align}\label{eq:jump_eq}
    [A] = r_s'(t){[u]}. 
\end{align}
This also holds for systems of quasi-linear equations, where $A,J$ and $u$ are vectors. However, the equations of motion in \eqref{eq:naive_set} give a shock location as 
\be 
r_s'(t) = \left[\int \mathrm{d}M N^x  \right]/[M],
\ee
which does not agree with the continuity of the first fundamental form in (\ref{eq:dynamical_junction}). To fix this, we change variables in \eqref{eq:naive_set}. This change keeps strong (classical) solutions the same, but it changes weak solutions. The discrepancy comes from the fact that replacing $u$ with $g=g(u)$ effectively multiplies \eqref{eq:1st_eq} by $g'(u)$. For smooth (strong) solutions, this scaling has no effect because $g'(u)$ is continuous. But for weak solutions, where $g'(u)$ may be discontinuous, the change introduces non-trivial physical effects that modify the structure of the solution. 

We now focus on the marginally bound case with $\epsilon = 0$ to simplify the equations and rewrite the system in the form of a balance equation. If we require the continuity of the first fundamental form \eqref{eq:dynamical_junction} as the jump condition, in principle we can obtain a matching form of the corresponding $A$ and variables $u$ in \eqref{eq:1st_eq} as follows. Since the variables in \eqref{eq:dynamical_junction} is $N^x$, we treat $u$ and $A$ as functions of $N^x$. This means we are changing variables from $N^x$ to $u$. Here we only focus on one variable $u$, since one equation is enough to determine the shock surface. Now, if we combine the jump condition \eqref{eq:jump_eq} with equation \eqref{eq:dynamical_junction}, we have 
\begin{align}\label{eq:junction_jump}
    \frac{A(N^x_+ , r) - A(N^x_- ,r)}{u(N^x_+) - u(N^x_-)} = -\frac{N^x_{-} + N^x_{+}}{2 } .
\end{align}
We want \eqref{eq:junction_jump} to hold for arbitrary values of $N^x_{\pm}$. In this sense, we can assume $N^x_{+} = N^x_{-} + \varepsilon$ and expand in powers of $\varepsilon$. This gives
\begin{align}
    0 =& (A' - u' N^x) \varepsilon + (u' + N^x u'' -A'') \varepsilon^2 \\
    &+ (3 u'' + 2 N^x u''' -2 A''') \varepsilon^3 + \mathcal{O}(\varepsilon^4).
\end{align}
By going through this process, we find that the only solution for $A$ and $u$ is
\begin{align}
    u = N^x, A = -\frac{(N^x)^2}{2},
\end{align}
up to an overall factor. This gives us back equation \eqref{eq:new_eom_Nx}.

Since multiplying the equations of motion by $r$ does not modify the weak solution(because $r$ is continuous), we define a new variable $X \equiv N^x r^l$, where $l$ is an arbitrary parameter. This transforms the equation of motion into
\be
\partial_t X  + \partial_r F(X, r) = J(X, r),\label{xdot} 
\ee
where the flux and source terms are
\be 
F(X, r) &= - \frac{X^2}{2 r^{l}},\\
J(X, r) = \frac{X^2(4-l)}{2 r^{l+1}}  &+ \frac{3 \left(\text{sgn}\sqrt{r^{2 l+2}-4 \zeta^2 X^2}-r^{l+1}\right)}{4 \zeta^2}. 
\ee 
The generalized velocity of the $X$ field is
\be 
v_X = - \frac{X}{r^l}. \label{eq:Xvelocity}
\ee
This follows the convention for a first order quasi-linear PDE of the form $\dot{u}=-v_u \partial_x u+J$, where $J$ is a source term that has no derivatives of $u$. In this form, the generalized velocity of the field $u$ is given by $v_u$. The equation \eqref{xdot} gives the dynamics, which we will study in the next sections.

\section{Numerical Algorithm to Solve Nonlinear PDEs}\label{Numerical results}
In this section, we consider the general problem of approximating solutions to the equation,
\be 
\partial_t X + \partial_r F(X,r) = J(X, r), \quad r\in \mathbb{R},\quad X\in \mathbb{R}, \label{eq:balance law}
\ee 
with the initial data given in (\ref{eq:InitialX}) and the boundary data in (\ref{eq:left bdry}) and (\ref{eq:right bdry}). The solution $X(t, r)$ propagates with a velocity $v_X$ (see (\ref{eq:Xvelocity})) that varies over space and time. 

The process of converting a continuous function into a finite set of values that can be stored on a computer is called \textit{discretization}. We divide the domain into a grid and store the function values at each grid point (see FIG. \ref{fig:grids}(a)). We write these values as $X_i^{n}=X(t^n,r_i)$, where $j$ is the position and $n$ is the time step. In the next subsections, we describe the numerical methods used for spatial and temporal discretization.

\subsection{Space update}
We discretize the field on a grid with $N$ points in the spatial direction, labeled $0, 1, \dots, N$. To update the values at the boundary points, $X^n_0 = X(t^n, r_{\mathrm{min}})$ and $X^n_N = X(t^n, r_{\mathrm{max}})$, we add ghost points $X_{-2}, X_{-1}, X_{N+1}$, and $X_{N+2}$, as shown in FIG. \ref{fig:grids}(a). We show how to update the solution in time in Section \ref{sec:time update}. In this part, to keep the notation simple, we drop the $t^n$ label.

\subsubsection{Central-Upwind Schemes}

We can approximate the spatial derivative using one-sided differences as either:
\be
\begin{aligned}
& \left(\partial_r X\right)_i \approx \frac{X_i-X_{i-1}}{\Delta r}+\mathcal{O}(\Delta r), \\
& \left(\partial_r X\right)_i \approx \frac{X_{i+1}-X_i}{\Delta r}+\mathcal{O}(\Delta r).
\end{aligned}
\ee 
where $X_i=X(r_i)$. We choose the upwind difference based on the direction of information flow. If $v_X > 0$, we use values to the left of point $i$. But the error in this method is proportional to $\Delta r$, so we often use a centered difference instead. A centered difference scheme uses values on both sides of the point to give a more balanced result. At grid point $r_i$, it is written as
\be
\left(\partial_r X\right)_i \approx \frac{X_{i+1} - X_{i-1}}{2 \Delta r} + \mathcal{O}(\Delta r^2),
\ee 
where $\mathcal{O}(\Delta r^2)$ shows that the error decreases faster as $\Delta r$ gets smaller. This centered difference typically provides a better approximation of the spatial derivative because it balances the information from both sides of the point. But when the solution has discontinuities or shocks, we will use the finite-volume method with the minmod slope limiter, which can better capture the flow of information in such cases.

\subsubsection{Finite-Volume Method with Minmod Slope Limiter}
\begin{figure}
    \centering
    \includegraphics[width=1.0\linewidth]{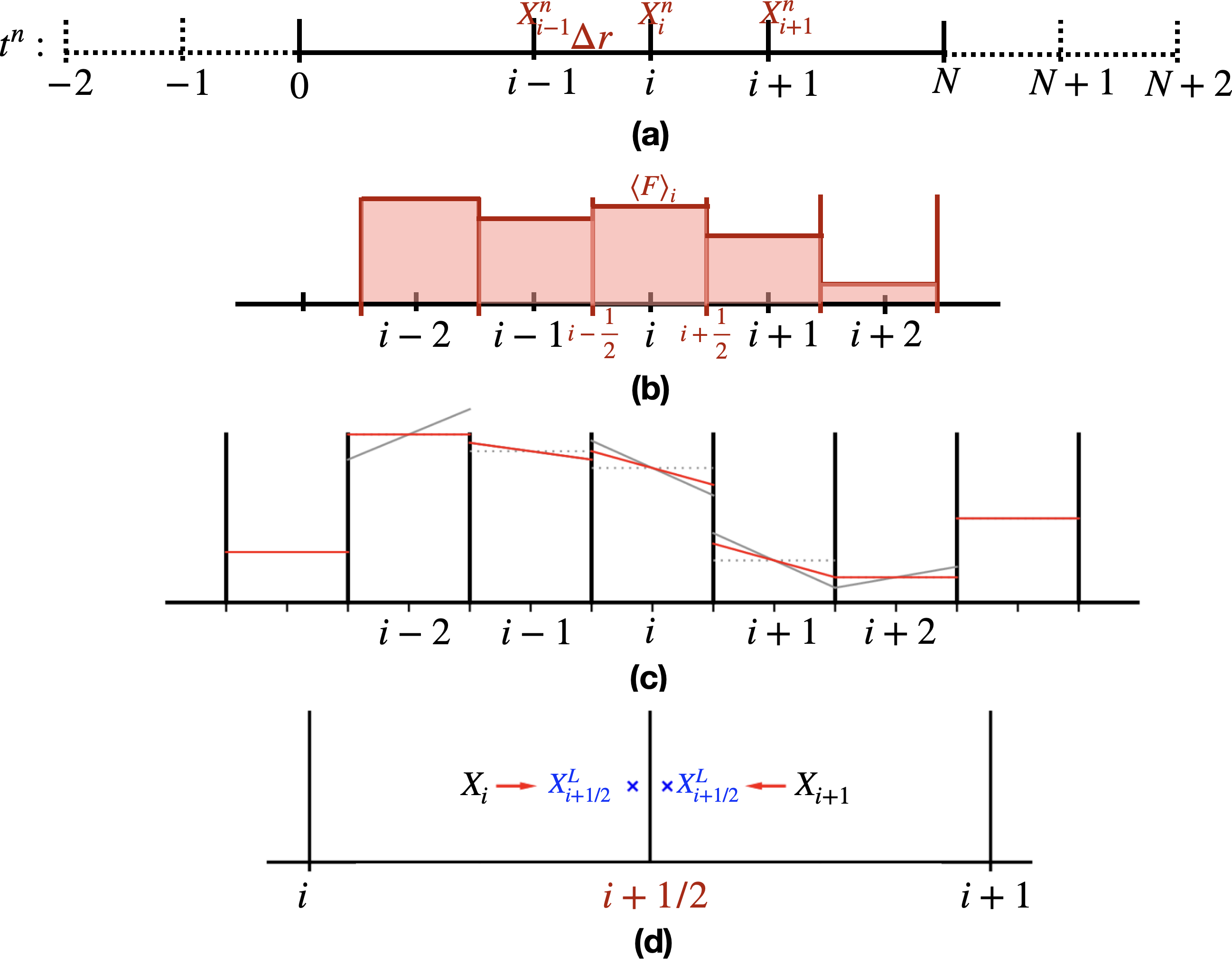}
    \caption{(a). Grids for spatial discretization. (b). Cells with the interfaces at the half grids for finite volume method. (c). Minmod slope limiter visualization. (d). Interface states. }
    \label{fig:grids}
\end{figure}

We solve the conservation law, 
\be 
\partial_t X + \partial_r F(X,r) = 0, \label{eq: Conservation Law}
\ee 
using a finite-volume method. Instead of storing data at specific grid points, we track the total amount (or average) inside each cell. The shaded region of a cell in \ref{fig:grids}(b) shows this concept. We take the average of equation (\ref{eq: Conservation Law}) over a cell $\left[r_{i-1 / 2}, r_{i+1 / 2}\right]$:
\be
\frac{1}{\Delta r} \int_{r_{i-1 / 2}}^{r_{i+1 / 2}} \frac{\mathrm{d} X}{\mathrm{d} t} \mathrm{d} r=-\frac{1}{\Delta r} \int_{r_{i-1 / 2}}^{r_{i+1 / 2}}\frac{\partial F}{\partial r} \mathrm{d} r. 
\ee 
Using the divergence theorem, we get the time derivative of the average value in a cell:
\be
\frac{\mathrm{d}}{\mathrm{d} t}\langle X\rangle_i=-\frac{1}{\Delta r}\left[F_{i+1 / 2}-F_{i-1 / 2}\right]. 
\ee
where the average in the cell is defined as 
\be 
\langle X \rangle_i(t):=\frac{1}{\Delta r} \int_{r_{i-\frac{1}{2}}}^{r_{i+\frac{1}{2}}} X(r, t) \mathrm{d} r.
\ee 
We compute the flux at the interface using the flux function:
$F_{i+1/2} = F(X_{i+1/2})$.  To approximate the flux at the interface, we need an estimate of $X$ at $r_{i + 1/2}$. For higher-order accuracy, we use piecewise linear reconstruction. This means we assume that $X(r)$ changes linearly inside each cell:
\be 
X(r) = \frac{\Delta X_i}{\Delta r}\left(r - r_i\right) + X_i,
\ee 
where the slope $\Delta X_i$ is computed using the minmod slope limiter \cite{MinmodLimiter} to ensure monotonicity and second-order accuracy:
\be
\Delta X_i=\operatorname{minmod}\left(X_i-X_{i-1}, X_{i+1}-X_i\right),
\ee
where the minmod limiter is defined as
\be
\operatorname{minmod}(a, b)= \begin{cases}a & \text { if }|a|<|b| \text { and } a \cdot b>0 \\ b & \text { if }|b|<|a| \text { and } a \cdot b>0 \\ 0 & \text { otherwise }\end{cases}.
\ee
The minmod limiter ensures that we only keep slopes where the left and right derivatives have the same sign. This helps avoid creating new peaks or dips. FIG. \ref{fig:grids} (c) shows how the limiter works: the gray lines show the original(unlimited) slopes, and the red lines show the slopes after applying the limiter. Notice that the slope at $i-1$ goes well above the original data on the right edge of the domain. That zone is a maximum. For zone $i-1$ the result of the limiting is to completely flatten the profile—we go to piecewise constant in that zone. 

\subsubsection{Riemann Problem and Shocks}
When reconstructing the solution, we obtain two values at each interface: one from the left and one from the right. These interface states are given by
\be 
X_{i+1/2}^{L} = X(r_{i}+\Delta r/2) = X_i + \Delta X_i/2,\\
X_{i+1/2}^{R} = X(r_{i+1}-\Delta r/2) = X_{i+1} - \Delta X_{i+1}/2,
\ee
see FIG. \ref{fig:grids}(d). To resolve the resulting degeneracy, we solve the \textit{Riemann problem} \cite{Rosini2013}. In the absence of shocks or rarefactions, the correct interface state is chosen as
\be
\begin{aligned}
X_{i+1/2}&=\mathcal{R}\left(X_{i+1 / 2}^L, X_{i+1 / 2}^R\right) \\
&= \begin{cases}X_{i+1 / 2}^L, & \text{for }v_{i+1 / 2}^L, v_{i+1 / 2}^R>0 \\ X_{i+1 / 2}^R, & \text{for }v_{i+1 / 2}^L, v_{i+1 / 2}^R<0\end{cases}\\
\end{aligned}
\ee
where $\mathcal{R}\left(X_{i+1 / 2}^L, X_{i+1/2}^ R\right)$ denotes the Riemann solution, and $v^{L(R)}_{i+1/2}=v_X(X^{L(R)}_{1+1/2})$ is the velocity of the field at the interface coming from the left (or right) grid. 

When information “piles up”  (i.e.,  $v_{i+1 / 2}^L > v_{i+1 / 2}^R$), which indicates intersecting characteristic curves, a shock forms. In that case, the shock speed is calculated using the Rankine-Hugoniot (jump) condition:
\be
v_s = \frac{F(X^{R}_{i+1/2})-F(X^{L}_{i+1/2})}{X^{R}_{i+1/2}-X^{L}_{i+1/2}}.
\ee 
Then, the solution to the Riemann problem is given by
\be
X_{i+1 / 2} =\left\{\begin{array}{cc}
X_s\, &\text{for } v_{i+1 / 2}^L> v_s >v_{i+1 / 2}^R \\
X_{\mathrm{other}}\, &\text{otherwise} 
\end{array}\right., 
\ee
where the shock state $X_s$ is defined as
\be
X_s= \begin{cases}X_{i+1 / 2}^L & \text { if } v_s>0 \\ X_{i+1 / 2}^R & \text { if } v_s<0\end{cases}
\ee
and the alternative state $X_{\mathrm{other}}$ is given by
\be
X_{\mathrm{other}}=\left\{\begin{array}{cc}
X_{i+1 / 2}^L & \text { if } v_{i+1 / 2}^L>0 \\
X_{i+1 / 2}^R & \text { if } v_{i+1 / 2}^R<0 \\
0 & \text { otherwise }
\end{array}\right.
\ee
Once the interface state $X_{i+1 / 2}$ is determined, the flux $F_{i+1/2}$ can be computed accordingly. 

While the Riemann solver is effective in capturing shock dynamics, alternative methods have been developed to reduce numerical viscosity and improve overall accuracy. One such method is the Kurganov and Tadmor (KT) central scheme, which avoids the need for an explicit Riemann solver while still achieving second-order spatial accuracy. 

\subsubsection{Kurganov and Tadmor (KT) Central Scheme}
KT central scheme \cite{KURGANOV2000241} is a finite-volume method that provides a Riemann-solver-free approach with second-order spatial accuracy. It is fully discrete and straightforward to implement. The scheme is expressed as
\be
\frac{\mathrm{d}}{\mathrm{d} t}\langle X\rangle_i =-\frac{1}{\Delta r}\left[F_{i+\frac{1}{2}}^*-F_{i-\frac{1}{2}}^*\right], \label{eq:KT}
\ee
where the central difference approximation for the flux $F_{i+\frac{1}{2}}^*$ is given by
\begin{equation}\footnotesize
\begin{aligned}
F_{i+\frac{1}{2}}^*=\frac{1}{2}\left[F\left(X_{i+\frac{1}{2}}^R\right)+F\left(X_{i+\frac{1}{2}}^L\right)-v_{i+\frac{1}{2}}\left(X_{i+\frac{1}{2}}^R-X_{i+\frac{1}{2}}^L\right)\right].
\end{aligned}    
\end{equation}
The local propagation speed $v_{i + \frac{1}{2}}$ is defined as
\be
\begin{aligned}
v_{i+\frac{1}{2}} = \max \left[\left|v_X\left(X_{i+1 / 2}^L\right)\right|, \left|v_X\left(X_{i+1 / 2}^R\right)\right|\right]    
\end{aligned}
\ee
This scheme introduces less numerical viscosity and can be implemented as either a fully discrete or semi-discrete scheme. Here we consider the semi-discrete scheme. With the KT scheme, one can verify that the right-hand side of (\ref{eq:KT}) is approximated on the $i$-th grid with second-order spatial accuracy as
\be 
\left(\frac{1}{2} X_i^2 + r_i X_i X_i'\right)
+\Delta r^2 \left(\frac{r_i}{6}  X_i X_i^{(3)}+\frac{1}{4} X_i X_i''\right.\\
\left.+\frac{1}{8} (X_i')^2+\frac{r_i}{4}  X_i' X_i''\right)+\mathcal{O}\left(\Delta r^3\right)
\ee 
where $X_i = X(r_i)$ and $X_i^{(n)} = X^{(n)}(r_i)$. In our numerical computation, we combine the Riemann solver and the KT central scheme. The Riemann solver is used at locations where shocks occur, and the KT central scheme is applied in regions with strong (classical) solutions. This combination ensures second-order accuracy for the strong solution over most of the spatial grid while reducing oscillations near shock locations.

\subsection{Discretization of the Source Term}
Averaging the balance law (\ref{eq:balance law}) over a cell $[r_{i-1/2}, r_{i+1/2}]$ yields
\be 
\frac{\mathrm{d}}{\mathrm{~d} t} \langle X\rangle_i=-\frac{F_{i+1/2}-F_{i-1/2}}{\Delta r}+\langle J\rangle_i. \label{DiscreteSource}
\ee
where the cell average of the source term $J(X(t,r), r)$ is defined as
\be 
\langle J \rangle_i:=\frac{1}{\Delta r} \int_{r_{i-\frac{1}{2}}}^{r_{i+\frac{1}{2}}} J(r, t) \mathrm{d} r.
\ee 
A semi-discrete scheme is obtained by applying an appropriate quadrature rule to approximate the spatial integral on the right-hand side of (\ref{DiscreteSource}) and by approximating the fluxes at the half-integer grid points $r = r_{i\pm 1/2}$. The method for approximating these fluxes was described in the previous section.

To approximate the cell-average of the source term while preserving second-order accuracy, we approximate $\langle S \rangle_i$ in the $i$-th cell as
\be 
\langle J\rangle_i = \frac{1}{2}\left[J(X^{L}_{i+1/2}, r_{i+1/2}) + J(X^{R}_{i-1/2}, r_{i-1/2})\right].
\ee
One can verify that this discretization achieves second-order accuracy, since the right-hand side expands to
\begin{multline}
\frac{5}{2} X_j^2 + \frac{3\Bigl(\text{sgn}\sqrt{1-4 \zeta^2 X_j^2}-1\Bigr)}{4 \zeta^2}   \\
\quad +\Delta r^2\left[\frac{5\,(X_j')^2}{8} - \frac{3\,(X_j')^2}{8}\,\text{sgn}\Bigl(1-4 \zeta^2 X_j^2\Bigr)^{-3/2}\right]
+ O\left(\Delta r^3\right)\nonumber
\end{multline}
for the special case $l = -1$.

A big challenge in discretizing this PDE is to determine the sign ($\mathrm{sgn}$) in the source term at each grid point during evolution. The first condition to change $\mathrm{sgn}$ happens when the value under the square root crosses zero, i.e., when $X_i \to \frac{r_i^{l+1}}{2\zeta}$. However, if a shock forms at $r_i$, then the sign of the source term at $r_i$ after the shock depends on the sign at its neighboring points $r_{i-1}$ or $r_{i+1}$:
\begin{itemize}
    \item If the shock is moving outward, the sign at $r_i$ is determined by the sign at $r_{i-1}$.
    \item If the shock is moving inward, the sign at $r_i$ is determined by the sign at $r_{i+1}$.
\end{itemize}
This neighbor-based rule ensures that the shock propagates and the sign updates consistently during the evolution.

\subsection{Time update} \label{sec:time update}
To achieve second-order accuracy in time integration, we construct second-order interface states in space and then use a second-order Runge–Kutta integrator to update the solution. This method-of-lines approach \cite{schiesser2009compendium} proceeds as follows:
\be
X_i^{(0)} &= X_i^{n},\\
X_i^{(1)} &= X_i^{(0)} + \frac{1}{2} R^j\left(X_i^{(0)}\right)\Delta t,\\
X_i^{(n+1)} &= X_i^{(0)} + R^j\left(X_i^{(1)}\right)\Delta t,   \label{eq:t updata}
\ee 
where $R^j$ is the spatial operator representing the right-hand side of (\ref{DiscreteSource}). The time step $\Delta t$ is chosen adaptively at each iteration according to
\be 
\Delta t = C \frac{\Delta r}{v_\mathrm{max}}
\ee 
with $v_\mathrm{max}$ being the maximum propagation speed at any cell interface and $C = 0.5$ the Courant–Friedrichs–Lewy (CFL) number\cite{1928MatAn.100...32C}. The CFL number is dimensionless, essential for ensuring convergence and measures the fraction of a cell traversed per time step. Equation (\ref{eq:t updata}) is an explicit update since its right-hand side depends solely on information from the previous time step.

\section{Implementation}\label{Implementation}
Let us apply these methods to our gravitational field problems. Specifically, we choose the special case $l=-1$ in (\ref{xdot}); extending the analysis to other values of $l$ is straightforward. We work in units where $G=1$.

\subsection{Initial Data}

\begin{figure}
    \centering
    \includegraphics[width=0.9\linewidth]{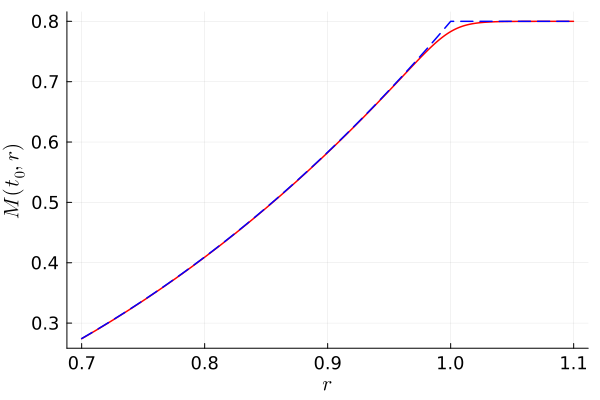}
    \caption{The effective dust mass  ${M}$  as a function of  $r$. The red curve shows the effective dust mass computed using the energy density given by (\ref{eq:density}). The blue dashed curve corresponds to the Oppenheimer-Snyder model \cite{Lewandowski:2022zce}, where  ${M} = \frac{4}{3} \pi r^3$  for  $0 \leq r \leq r_0$  and remains constant at  $M_0=0.8$  for  $r > r_0$. }
    \label{fig:mass_original}
\end{figure}

Before we investigate the dynamics of the $X$ field in (\ref{xdot}), we adopt the following initial data $X(t_0, r)$ and boundary data $X(t, r_{\mathrm{min}(\mathrm{max})})$.
We use a smooth step function to describe the initial energy density of the dust field:
\be 
\rho(t_0, r) = \rho_0 \frac{e^{k (r_0 - r)}}{1 + e^{k (r_0 - r)}}. \label{eq:density}
\ee
Here,  $k \geq 0$ controls the steepness of the transition. We set  $k = 100$  and  $r_0 = 1$  to create a smooth step function transitioning from $0$ to $\rho_0 = \frac{3M_0}{4\pi r_0^3}$, centered at  $r_0 = 1$, with total mass $M_0 = 0.8$. The dust mass function is then given by
\be
M(t, r) = 4\pi \int_0^r \mathrm{d}\tilde{r} \tilde{r}^2 \rho(t, \tilde{r}). \label{eq:eff mass}
\ee
Using the initial energy density in (\ref{eq:density}), we compute the initial mass profile  ${M}(t_0, r)$, as shown in FIG. \ref{fig:mass_original}. Unlike the Oppenheimer-Snyder model (see blue dashed curve in FIG. \ref{fig:mass_original}), where the mass sharply transitions to $M_0 = 0.8$, the effective mass smoothly increases from $0$ to $0.8$.

\begin{figure}
    \centering
    \includegraphics[width=0.9\linewidth]{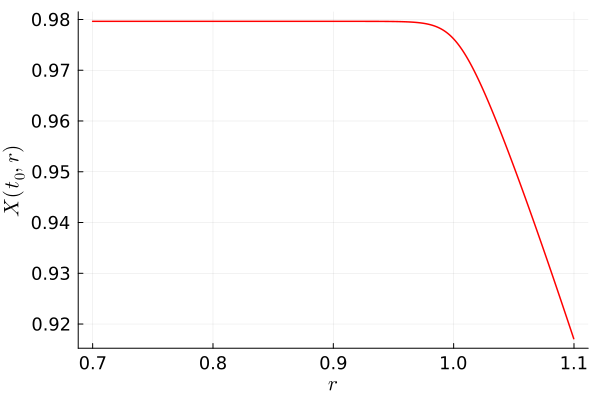}
    \caption{The initial condition for  $X(t_0, r)$ in (\ref{eq:InitialX}), derived from the effective mass function in (\ref{eq:eff mass}) with the quantum parameter $\zeta=0.5$ and $l=-1$.}
    \label{fig:X_initial}
\end{figure}

By setting $\epsilon = 0$ in \eqref{eq:nx_in_m}, the dust mass is then expressed in terms of the $X$ field as
\be 
{M} = \frac{r^2 \left(r+ \text{sgn} \sqrt{r^{2}-4\zeta^2 r^{-2l} X^2}\right)}{4 \zeta^2}. 
\ee 
From this, we determine the initial condition for $X(t_0,r)$:
\be
X(t_0, r) = \sqrt{2{M}r^{2l-1}-4M^2\zeta^2r^{2l-4}}. \label{eq:InitialX}
\ee 
This is shown in FIG. \ref{fig:X_initial}. Here, we choose the positive root so that the profile initially represents a collapsing configuration rather than an expanding one. 

\subsection{Boundary Data}\label{Sec:Bdry data}
Boundary conditions are imposed at the boundaries of the domain $r\in[0.7, 1.1]$. The inner boundary data $X(t, r_{\mathrm{min}})$ and outter boundary data  $X(t, r_{\mathrm{max}})$ can be determined from the analytical solution to the equation of motion in (\ref{xdot}).

\begin{figure}
    \centering
    \includegraphics[width=1.0\linewidth]{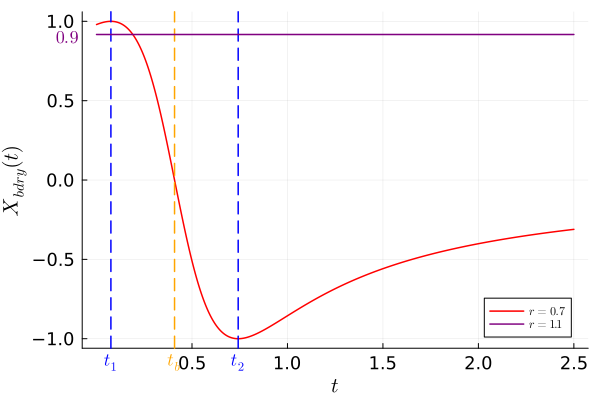}
    \caption{Boundary data for $X_{\text{bdry}}(t)$ at the left $(r_{\mathrm{min}}=0.7)$ and right $(r_{\mathrm{max}}=1.1)$ boundaries. The red curve corresponds to the evolution of the solution in the dust region at $r_{\mathrm{min}}$, and the purple horizontal line shows the constant solution in the vacuum region at $r_{\mathrm{max}}$. The dashed blue lines mark the times $t_1 \simeq 0.075$ and $t_2 \simeq 0.742$, where (\ref{xdot}) changes sign, and the dashed orange line at $t_b \simeq 0.409$ indicates the bounce time. The label $X(t)=0.9$ highlights the constant value in the vacuum region at $r_{\mathrm{max}}$.}
    \label{fig:Xbdry}
\end{figure}

\textbf{Analytical solution for} $r_\mathrm{min}\ll r_0$ (i.e., the boundary of the homogeneous dust region): Since the interior is a portion of an FLRW spacetime, the solution  $X(t, r)$ is proportional to the scale factor $A(t)$ of the FLRW spacetime. Let us introduce the ansatz:
\be
X(t,r) = A(t) r^{l+1}.
\ee 
Substituting this ansatz into the equation of motion (\ref{xdot}) yields an ODE whose solution is
\be 
A(t) = \frac{-6 t-8 c_1 \zeta ^2 r^{l+1}}{4 \zeta ^2+\left(3 t+4 c_1 \zeta ^2 r^{l+1}\right){}^2}.
\ee 
Thus, we obtain
\be 
X(t, r) = -\frac{2\left(3 t+4 c_1 \zeta ^2 r^{l+1}\right)r^{l+1}}{4 \zeta ^2+\left(3 t+4 c_1 \zeta ^2 r^{l+1}\right)^2}. \label{eq:left bdry}
\ee
According to the initial data in (\ref{eq:InitialX}), taking $t_0 = 0$ and, for example, $l = -1$, we obtain 
\be 
c_1 \simeq -1.226,
\ee 
which is a constant for the dust region boundary solution. In the numerical computation, the left boundary  $r = r_{\mathrm{min}} = 0.7$ is always in the homogeneous region; therefore, we can set the left boundary data using (\ref{eq:left bdry}). According to the analytical result for the inner solution, the sign in (\ref{xdot}) changes from  $1$  to $-1$ and from $-1$ to  $1$  when  
\be 
X(t,r) = \pm \frac{r^{l+1}}{2\zeta}
\ee
at  $t_{1}$  and  $t_{2}$ (see FIG. \ref{fig:Xbdry}). The bounce occurs at
\be
t_b = \frac{t_1 + t_2}{2}
\ee 
when $X = 0$.

\textbf{Analytical solution for} $r_{\mathrm{max}}\gg r_0$: In the vacuum region, the solution is given by the Schwarzschild metric. Here, we introduce the ansatz
\be
X(t,r) = B(r),
\ee 
which is independent of time. For convenience, we consider the special case $l = -1$. Then the solution to (\ref{xdot}) is
\be 
X(t, r) = \frac{\sqrt{-c_2 \left(c_2-2 r^3\right)}}{r^3}. \label{eq:right bdry}
\ee 
From the initial data in (\ref{eq:InitialX}), we obtain
\be 
c_2 \simeq 0.800,
\ee 
which is a constant in the vacuum region solution. The bounce occurs at
\be
r_b \simeq 0.737,
\ee 
where $X = 0$. This also indicates that the chosen minimum radius, $r_{\mathrm{min}} < r_b$, indeed lies entirely within the homogeneous dust region.

\subsection{Method of characteristics} \label{Sec:Method of characteristics}
The method of characteristics is a technique for solving first-order PDEs by reducing them to a family of ODEs \cite{evans10}. The characteristic equations for (\ref{xdot}) are
\be
\begin{cases}
    \frac{\rmd r}{\rmd t}= - \frac{X}{r^l} = - N^x \\
    \frac{\rmd X}{\rmd t}= \frac{(2-l)X^2}{r^{1 + l}} 
+ \frac{3\left(-r^{l+1}+ \text{sgn} \sqrt{r^{2 + 2l} - 4 \zeta^2 X^2}\right)}{4 \zeta^2} \label{Charateristics}
\end{cases} 
\ee
These equations describe the evolution of $X(t,r)$ along the trajectories $r(t)$. When the characteristic curves do not intersect, each trajectory yields a \textit{strong (classical) solution} of the PDE.

\begin{figure}
    \centering
    \includegraphics[width=1.0\linewidth]{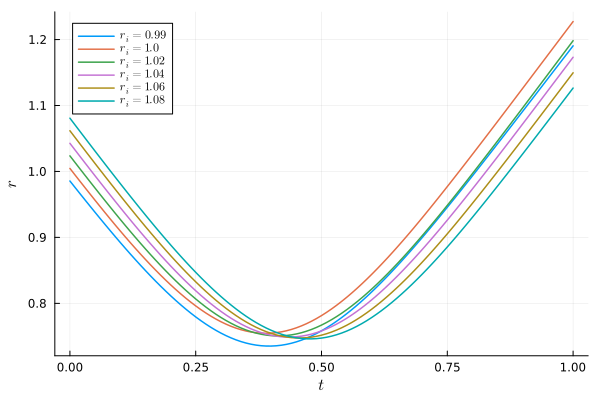}
    \caption{Characteristic curves of $r(t)$ as described by (\ref{Charateristics}) for different initial positions $r_i$. Each color corresponds to a distinct trajectory starting from a different $r_i \in \{0.99, 1.00, 1.02, 1.04, 1.06, 1.08\}$, illustrating how the radius $r$ evolves over time for each initial position. Because of the specific initial data chosen, the characteristic curves intersect at certain points.}
    \label{fig:charateristics_original}
\end{figure}

However, with the initial data in (\ref{eq:InitialX}), the characteristic curves (see FIG. \ref{fig:charateristics_original}) intersect at certain points. This intersection indicates that a strong solution no longer exists, and the $X$ field develops a discontinuity for $t$; such a moving discontinuity is called a \textit{shock wave}, and we denote its position by $r_s(t)$. In such cases, a \textit{weak solution} is required. In gravitational fields, weak solutions are essential for capturing dynamics involving shock formation \cite{Fazzini:2023ova}. 

\subsection{Numerical weak solution}

\begin{figure}
    \centering
    \includegraphics[width=0.9\linewidth]{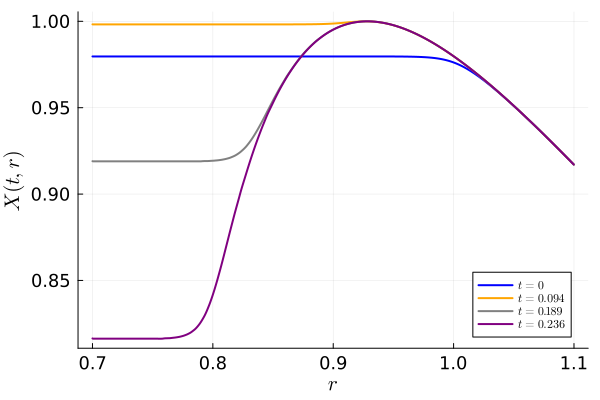}
    \caption{Early evolution: The solution $X(t,r)$ remains continuous. The curves show how the solution changes smoothly from $r_{\mathrm{min}} = 0.7$ to $r_{\mathrm{max}} = 1.1$ at different times before the bounce.}\label{fig:Xvsr_difft1}
\end{figure}

\begin{figure}
    \centering
\includegraphics[width=0.9\linewidth]{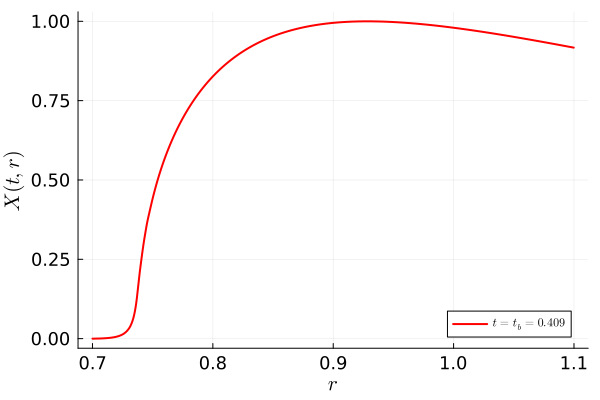}
    \caption{At the bounce moment $t_b$: The red curve shows the system at its maximum compression (the bounce). Beyond this point, the evolution transitions from collapse to expansion.}\label{fig:bounce}
\end{figure}

\begin{figure}
    \centering
\includegraphics[width=0.9\linewidth]{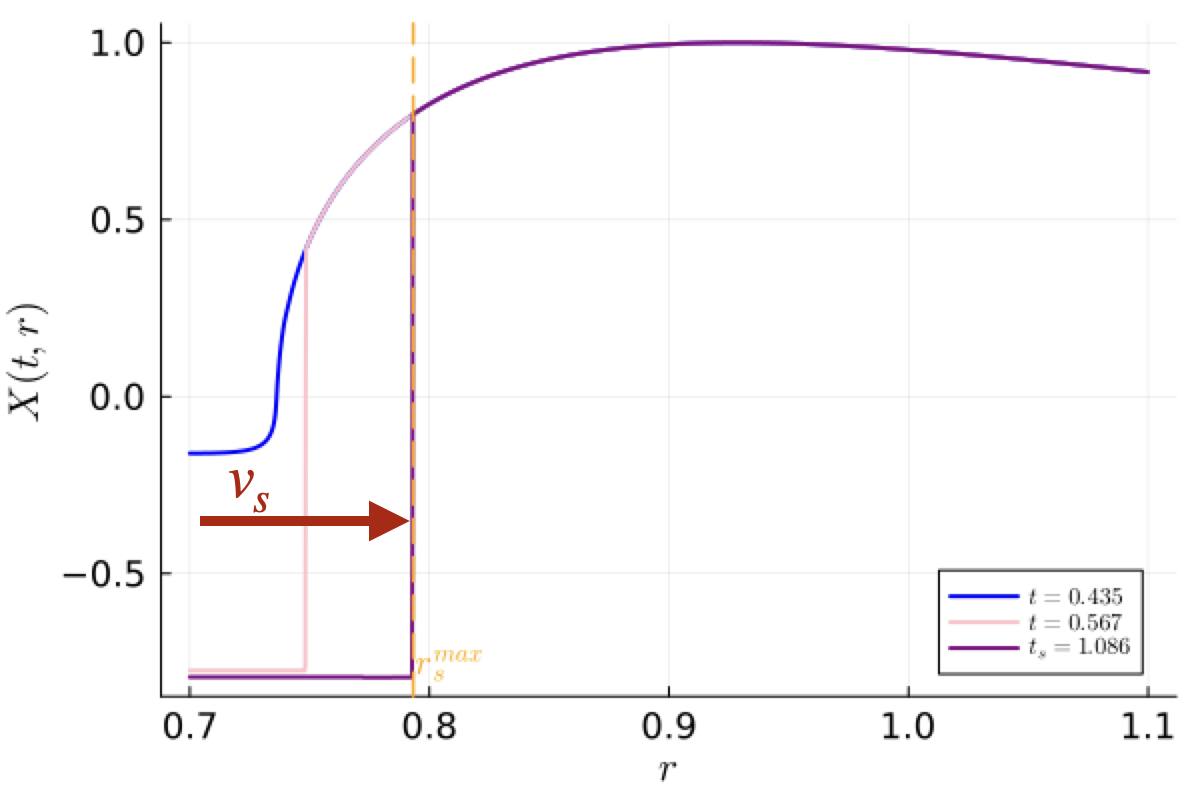}\caption{Post-bounce shock formation: A discontinuity appears in the curves, indicating the formation of a shock. The arrow labeled $v_s$ shows the shock propagating outward.}\label{fig:Xvsr_difft3}
\end{figure}

\begin{figure}
    \centering
\includegraphics[width=0.9\linewidth]{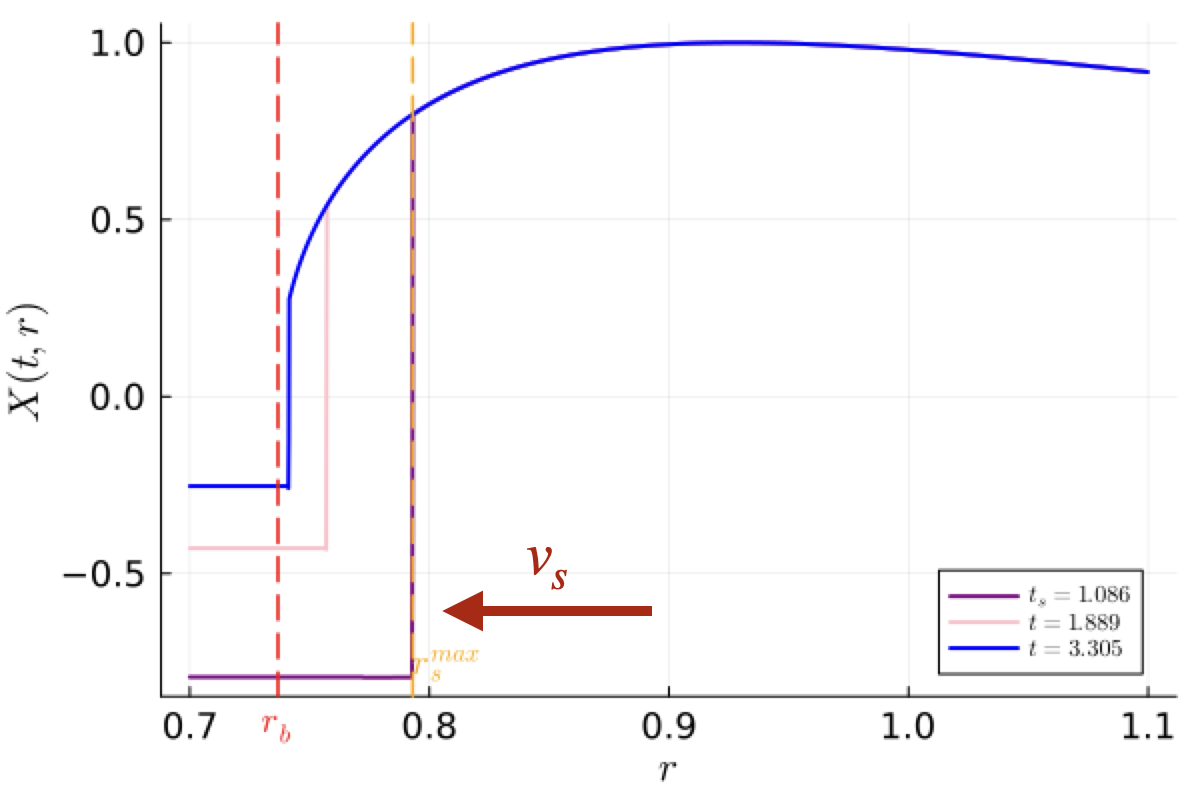}
    \caption{Shock reversal and inward motion: After reaching its maximum radius $r_s^{\mathrm{max}}$, the shock begins moving inward. The dashed red line marks the bounce location $r_b$, and the arrow indicates the shock velocity $v_s$ directed toward $r_b$; however, the shock never reaches $r_b$.}
    \label{fig:Xvsr_difft4}
\end{figure}

\begin{figure}
    \centering
    \includegraphics[width=1.0\linewidth]{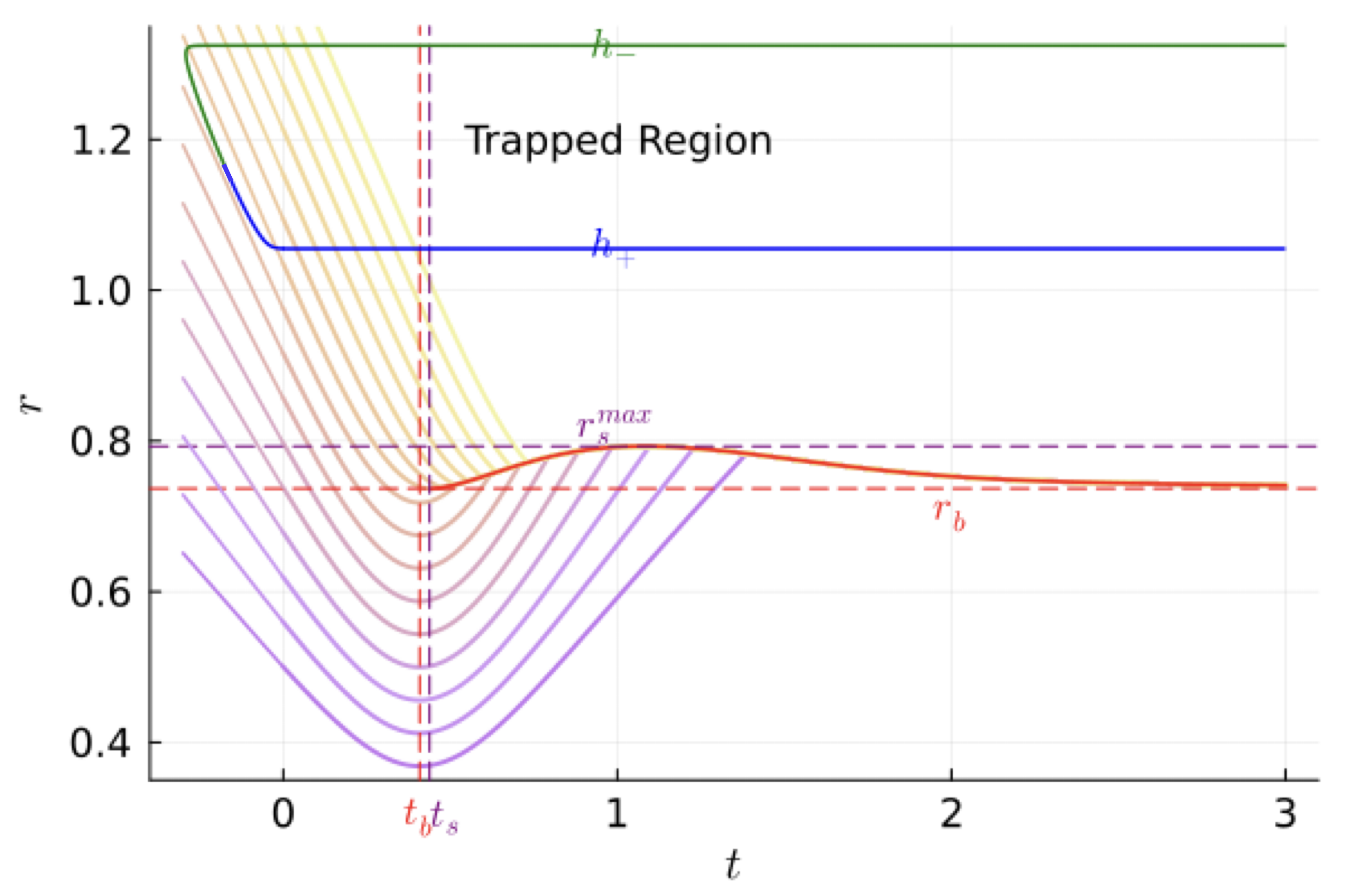}
    \caption{Characteristic curves $r(t)$ for different initial radii with weak solutions when $M^{\mathrm{max}}(t_0)=0.8$. A shock forms after the bounce at $t_s$ and moves toward the bounce location $r_b$. The blue and green curves show the inner horizon $h_+$ and the outer horizon $h_-$. These horizons are dynamical when $t$ is small but stay constants at $h_+ \to  1.055$ and  $h_-\to 1.325$ as time increase.}
    \label{fig:characteristic_shock}
\end{figure}

Now we use the numerical method given in Sec. \ref{Numerical results} to search for weak solutions.
FIG. \ref{fig:Xvsr_difft1} illustrates the early evolution of the star’s collapse, where the solution $X(t,r)$ remains continuous and smoothly transitions from $r_{\mathrm{min}} = 0.7$ to $r_{\mathrm{max}} = 1.1$ prior to the bounce. FIG. \ref{fig:bounce} captures the bounce moment at $t=t_b=0.409$: the red curve reaches its maximum compression, marking the transition from collapse to expansion. FIG. \ref{fig:Xvsr_difft3} then shows the formation of a shock shortly after the bounce, at $t_s = 0.435$. Here, a discontinuity appears in the curves, and the shock surface initially propagates outward until it reaches its maximum position (the orange dashed line) at $r_s^{\mathrm{max}}=0.793$. Finally, FIG. \ref{fig:Xvsr_difft4} indicates the shock reversal and inward motion: after reaching its maximum radius $r_s^{\mathrm{max}}$ at $t_s = 1.086$, the shock surface subsequently moves inward and asymptotically approaches the minimal radius $r_b = 0.737$ of the vacuum solution with $M=0.8$ where bounce happens. Over time, the star's mass asymptotically concentrates within the shock wave, leading its interior (within the minimal radius) to approach an asymptotically flat geometry. This is validated by FIG. \ref{fig:characteristic_shock} and curvature invariants analyzed in Subsection \ref{Scalar curvatures}.

To investigate the characteristic curves of the system, we substitute the numerical solution $X(t,r)$ into the ODE:
\be
\frac{\mathrm{d}r}{\mathrm{d}t} = -r X.
\ee 
Solving this ODE for various initial conditions yields the trajectories $r(t)$ shown in FIG. \ref{fig:characteristic_shock}. Before shock formation, these characteristic curves indicate how $X(t,r)$ evolves smoothly in time and space. However, shortly after the bounce time $t_b$, a shock forms at $t_s$. Once the shock appears, the PDE solution has a weak solution at the shock surface, because the solution is no longer differentiable there. 

Numerically, we observe that all trajectories from  different starting radii eventually converge onto the shock surface. This shows that the shock front acts as an attractor for the characteristics. The red curve represents the shock position at different times 
\begin{equation}\small
\begin{aligned}
     r_{s}(t) = \left(1.31 - 4.22t + 12.41t^2 - 19.47t^3 + 19.04t^4 - 12.45t^5 \right.\\
 \left.+ 5.56t^6 - 1.67t^7 + 0.32t^8 - 0.04t^9 +  0.002t^{10}\right)_{\pm 4.74\times 10^{-4}},  \label{rshock}   
\end{aligned}
\end{equation}
which is given by interpolation to a polynomial functions with the error $\pm 4.74\times 10^{-4}$. 

\begin{figure}[ht!]
    \centering
    \includegraphics[width=1.0\linewidth]{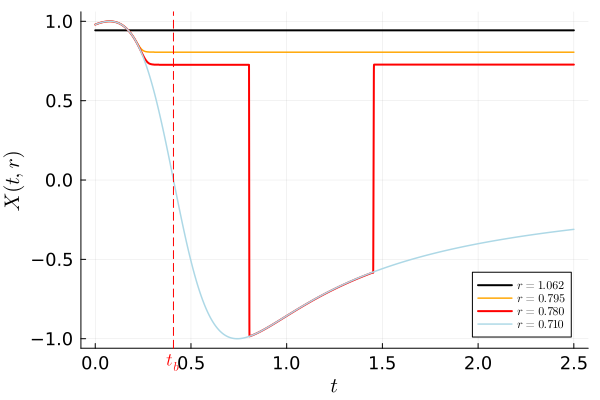}
    \caption{Time evolution of $X(t,r)$ at various fixed radii. Each color shows a different radial location, illustrating transitions between dust and vacuum regions as well as shock propagation.}
    \label{fig:Xvst_diffr}
\end{figure}

FIG. \ref{fig:Xvst_diffr} shows the evolution of $X(t,r)$ at several fixed radii $r_\mathrm{const}$ (these radii appear as horizontal lines in FIG. \ref{fig:characteristic_shock}). We identify four main regimes:
\begin{itemize}
    \item $r_\mathrm{const} > r_0$ (black line): This radius remains in the vacuum region throughout the entire evolution, so the solution follows the Schwarzschild solution in (\ref{eq:right bdry}).
    \item $r_s^{\mathrm{max}}<r_\mathrm{const} < r_0$ (orange line): This radius is initially within the dust region, then transitions to vacuum after some time. Once it is in the vacuum region, $X$ remains constant.
    \item $r_b<r_\mathrm{const} < r_s^{\mathrm{max}}$ (red line): This radius at first lies in the dust region, then transitions to the vacuum region. When the shock passes, it switches back to the dust. As the shock later reverses and crosses this radius again, the solution returns to vacuum solution, where it remains thereafter.
    \item $r< r_{b}$ (light-blue line): This radius stays in the dust region for the entire evolution. The solution follows (\ref{eq:left bdry}). 
\end{itemize}
The star has an initial dust boundary at $\sim r_0$, a maximum shock radius at $r_s^{\mathrm{max}}$, and a bounce radius for the vacuum at $r_b$. These four cases show how different parts of the star go through different phases—ranging from simple dust collapse to multiple transitions between dust and vacuum caused by shock propagation.

\subsection{Signature of the shock surface}\label{signature_of_shock}

\begin{figure}[ht!]
\centering
\includegraphics[width=1.0\linewidth]{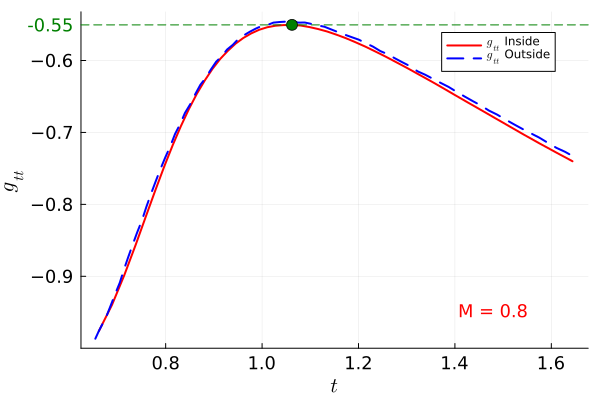}
\includegraphics[width=1.0\linewidth]{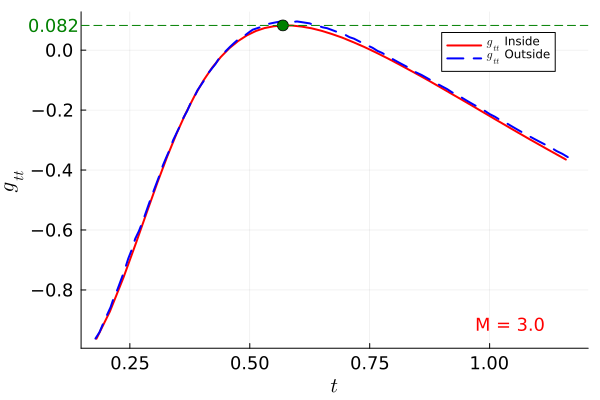}
\caption{The induced metric component $g_{tt}$ along the shock surface for two different effective mass values. The red and blue dashed curves show $g_{tt}$ as a function of time, computed from the inside and outside, respectively. \textbf{Top panel}: For $M = 0.8$, the entire shock surface remains timelike throughout the evolution, as indicated by $g_{tt} < 0$. \textbf{Bottom panel}: For $M = 3.0$, the induced metric becomes positive at its maximum, indicating that the shock surface becomes spacelike during part of the evolution. The green dashed lines mark the peak value of $g_{tt}$ computed from the inside, and the green dot highlights the maximum point.}\label{fig:gtt}
\end{figure}

By construction, the induced metric \eqref{eq:induced_metric} on the shock surface must be continuous. We verify this explicitly in FIG. \ref{fig:gtt}, which shows the $g_{tt}$ component of the induced metric on both sides of the shock surface. The red and blue dashed curves represent the values from the inside and outside, respectively, and they overlap closely throughout the evolution, confirming the continuity of the induced metric. Although the two curves appear slightly separated in FIG. \ref{fig:gtt}, this is due to spatial discretization and the use of piecewise linear plotting. The small gap results from numerical errors associated with the finite number of grid points. Increasing the resolution reduces this error and brings the curves into closer agreement.

With a continuous induced metric, we can determine the causal nature of the shock surface, specifically whether it is physical (timelike) or phantom (spacelike). The sign of \(g_{tt}\) indicates this causal structure: if \(g_{tt} < 0\), the surface is timelike; if \(g_{tt} > 0\), it is spacelike. As shown in FIG. \ref{fig:gtt}, for small effective masses (e.g., \(M = 0.8\)), the entire shock surface remains timelike throughout the evolution, meaning it represents a physical shock. However, increasing the effective mass changes this behavior. For example, when \(M = 3.0\) (lower panel of FIG. \ref{fig:gtt}), the induced metric component \(g_{tt}\) becomes positive in certain regions, indicating that parts of the shock surface become spacelike.

\begin{figure}[ht!]
\centering
\includegraphics[width=1.0\linewidth]{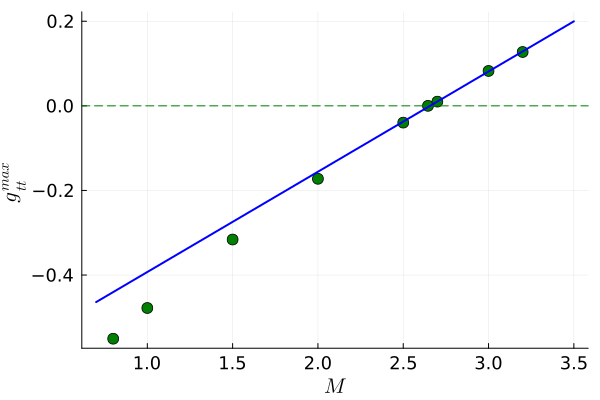}
\caption{Maximum value of the induced metric component $g_{tt}$ (computed from the inside) along the shock surface, plotted as a function of the effective mass $M$. The green dashed line at $g_{tt} = 0$ marks the threshold between timelike ($g_{tt} < 0$) and spacelike ($g_{tt} > 0$) regions. As $M$ increases, the maximum value of $g_{tt}$ shifts from negative to positive, indicating a change in the causal nature of the shock surface. The blue line is a linear fit to the last five data points, showing a clear linear trend in the large-mass region.}\label{fig:gttVsM}
\end{figure}

To systematically examine this transition, we plot the maximum value of \(g_{tt}\) as a function of the effective mass \(M\) in FIG. \ref{fig:gttVsM}. For small masses (\(M < 2.645\)), the maximum of \(g_{tt}\) stays negative, showing that the shock surface retains its physical (timelike) nature throughout. As \(M\) increases beyond approximately \(2.645\), the maximum \(g_{tt}\) crosses zero for the first time and becomes positive. This change marks the appearance of spacelike segments on the shock surface, meaning it becomes phantom in those regions. This transition shows that the effective mass plays a key role in determining the causal structure of the shock.

This observation suggests that physical shocks (timelike) are only possible for black holes with extremely small masses, such as those near Planck-scale masses. For larger black holes, the appearance of spacelike segments indicates limitations in the current effective theory when describing the full dynamics around shocks. In such cases, effects like quantum tunneling may become important and fall outside the scope of standard dynamics. Interestingly, in the large-mass regime, the trend in the maximum value of $g_{tt}$ appears to follow a linear pattern (See FIG. \ref{fig:gttVsM}). However, as black holes evaporate and their mass drops (e.g., approaching or falling below \(M \approx 2.645\)), the conditions for forming a physical shock are restored. This implies that to fully understand black hole dynamics and spacetime structure, evaporation must be taken into account in theoretical models.

\subsection{The formation of trapped and anti-trapped regions }\label{signature_of_shock}

The location of the horizon can be determined from the expansion parameter $\theta_{\pm}$ of the two future-directed null vector fields that are normal to a constant-radius shell. In LTB coordinates, the horizons are located at  $\dot{R} = \pm 1$ \cite{Giesel:2023hys}. In PG coordinates, this corresponds to $N^x = r X(t,r) = \mp 1$, according to Eq.\eqref{eq:nx_in_m}, which gives an inner horizon at $h_{+} \approx 1.325$ and an outer horizon at $h_{-} \approx 1.055$ in the vacuum region (see FIG. \ref{fig:characteristic_shock}).

The appearance of a trapped region implies the formation of a black hole. This has been shown to occur only when the mass exceeds a critical value \( M_c = \frac{8\zeta}{3\sqrt{3}G} \) \cite{Giesel:2023hys}. The total mass \( M_0 = 0.8 \) used in our numerical evaluation slightly exceeds this threshold (\( M_c \simeq 0.77 \) for \( \zeta = 0.5 \)).  The region between the inner and outer horizons forms a trapped zone. As shown in FIG. \ref{fig:characteristic_shock}, any dust geodesic that enters this trapped region is inevitably pulled toward the shock, before reaching its minimal radius where the bounce would otherwise occur.

\begin{figure}[ht!]
    \centering
    \includegraphics[width=1.0\linewidth]{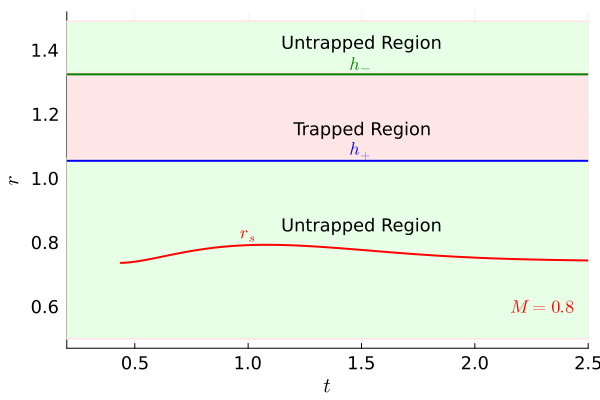}
    \includegraphics[width=1.0\linewidth]{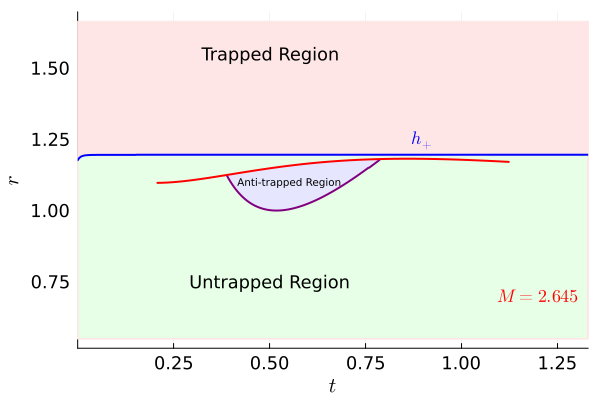}
    \includegraphics[width=1.0\linewidth]{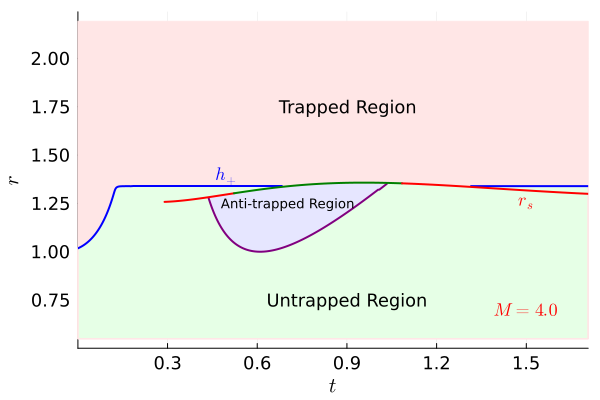}
    \caption{Spacetime region classification for different effective masses $M=M^{\mathrm{max}}(t_0)$. Each panel shows the evolution of the inner horizons $h_+$ (blue), the event horizon $h_-$ (green), and the shock surface $r_s$. The shock surface is color-coded: green for spacelike segments and red for timelike segments. The background shading highlights different causal regions: the trapped region (pink) lies between the inner and outer horizons, while the untrapped regions (light green) appear above $h_-$ and below $h_+$. For larger mass ($M = 2.645$ and $M = 4.0$), an anti-trapped region (bounded by $N^x=-1$ (shaded in purple) and shock surface) temporarily emerges. }
    \label{fig:regions_diffM}
\end{figure}

The expansion parameters help identify different causal regions of spacetime. A trapped surface (light pink shaded region in FIG. \ref{fig:regions_diffM}) forms when both $\theta_{
\pm} < 0$ \cite{Hayward:1994bu}. An anti-trapped region (light purple shaded region in FIG. \ref{fig:regions_diffM}) appears when $\theta_{\pm} > 0$; the purple curve in FIG. \ref{fig:regions_diffM} marks the outer horizon of the untrapped region, where $\theta_{-} = 0$. The remaining regions are untrapped (light green shaded region). As shown in FIG. \ref{fig:regions_diffM}, for small masses, the shock surface is timelike and lies entirely within the untrapped region, with no anti-trapped region present. When the masses increase but remain below $M \simeq 2.645$, the shock surface is still timelike, but an anti-trapped region begins to appear. This region is partially intersected by the shock surface during the gluing of different spacetime patches, although the shock surface itself remains inside the untrapped region. For large masses, the shock surface crosses the inner horizon of the trapped region, and parts of it become spacelike.

\subsection{Curvatures inside and outside along the shock surface}\label{Scalar curvatures}

\begin{figure}
\centering
\includegraphics[width=1.0\linewidth]{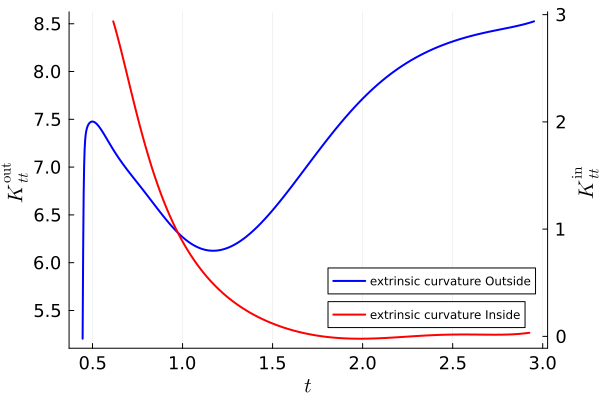}
\includegraphics[width=1.0\linewidth]{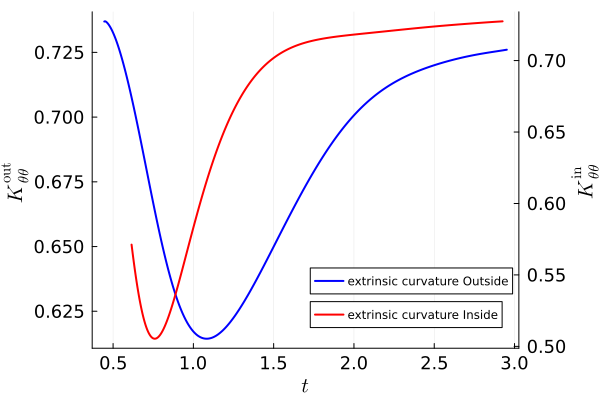}\label{fig:K_theta_theta}
\caption{Evolution of the extrinsic curvature component  $K_{tt}$ and $K_{\theta\theta}$ inside (red curve) and outside (blue curve) along the shock surface.}
\label{fig:3d_curvature}
\end{figure}

As we know, the induced metric describes the intrinsic geometry of the shock surface. However, it does not tell us how the surface is embedded in the surrounding spacetime. To understand this, we compute the extrinsic curvatures, which describes how the shock surface bends and evolves over time. The extrinsic curvature components are:
\be
    K_{tt} &= \frac{-{\partial_r N^x} \left({N^x}+\frac{\mathrm{d}r_s}{\mathrm{d}t}\right)^3+{N^x} {\partial_r N^x}-{\partial_t N^x}-\frac{\mathrm{d}^2r_s}{\mathrm{d}t^2} }{\sqrt{|1-\left(N^x+\frac{\mathrm{d}r_s}{\mathrm{d}t} \right)^2|}} \\
    K_{\theta \theta} &= -\frac{r_s \left(N^x \left(N^x+\frac{\mathrm{d}r_s}{\mathrm{d}t} \right)-1\right)}{\sqrt{|1-\left(N^x+\frac{\mathrm{d}r_s}{\mathrm{d}t} \right)^2|}}
\ee
FIG. \ref{fig:3d_curvature} shows both $K_{tt}$ and $K_{\theta \theta}$, , computed on both sides of the shock surface. Both components have discontinuities across the surface. 
This confirms that the junction condition does not satisfy the standard Israel junction condition in the absence of surface energy-momentum tensor, which would require the extrinsic curvature (second fundamental form) to be continuous across the junction surface. The mismatch arises from two main reasons. First, the shock surface naturally incorporates non-trivial matter, leading to a localized energy-momentum tensor contribution along the boundary—a phenomenon also appears in classical general relativity \cite{Tegai:2011qf,Husain:2025wrh}. Second, within effective dynamics, an alternative extrinsic curvature matching conditions is needed for junction surface that are not aligned with dust geodesics, as shown in \cite{Giesel:2023hys}. Consequently, the extrinsic curvature here is discontinuous at the shock surface. A comprehensive derivation and validation of these modified junction conditions and the corresponding surface energy-momentum tensor will be detailed in a subsequent paper.

It is also important to examine the behavior of curvature scalars. Since the extrinsic curvature is discontinuous, we expect the curvature scalars to be discontinuous as well, though still finite. For example, in PG coordinates, the Ricci curvature is given by:
\be
\mathcal{R} = \frac{2 \left(N^x\right)^2}{r^2} &+ \frac{8 N^x \partial_r N^x - 4\partial_t N^x}{r}\\
 + &2N^x\partial^2_{r}N^x +2\left(\partial_{r}N^x\right)^2-2 \partial_{t}\partial_r N^x. \label{eq:Ricci curvarute}
\ee
In the classical limit where  $N^x \rightarrow \sqrt{2M/r}$, the Ricci scalar approaches zero as expected.

\begin{figure}
    \centering
    \includegraphics[width=1.0\linewidth]{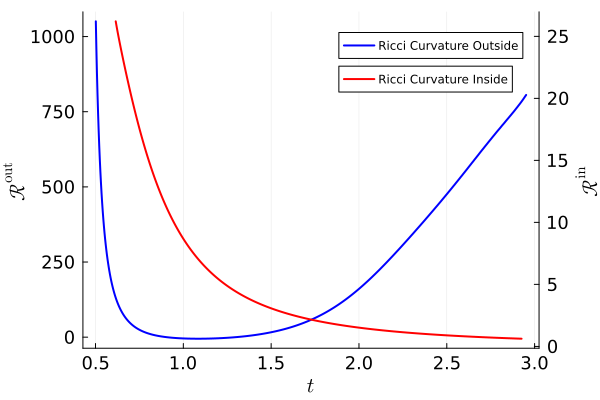}
    \caption{The Ricci curvature evolution outside (inside) along the shock surface. The blue (red) curve denotes the Ricci curvature outside (inside) along the shock  $\mathcal{R}^{\mathrm{out}}$($\mathcal{R}^{\mathrm{in}}$). The curvature is discontinuous at the shock surface, and  $\mathcal{R}^{\mathrm{in}}$ approaches zero as time increases.}\label{fig:Ricci_curvature}
    \includegraphics[width=1.0\linewidth]{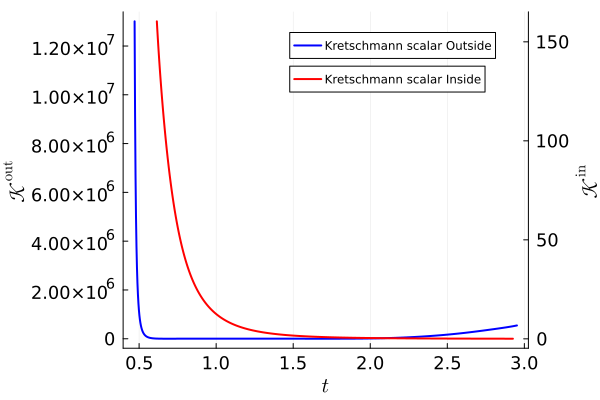}
    \caption{The Kretschmann scalar evolution outside (inside) along the shock surface. The blue (red) curve denotes the Kretschmann scaler outside (inside) along the shock  $\mathcal{K}^{\mathrm{out}}$($\mathcal{K}^{\mathrm{in}}$).}
    \label{fig:Kretschmann_curvature}
\end{figure}

To study the behavior of $\mathcal{R}$, we compute its values just inside and outside along the shock surface, denoted as $\mathcal{R}^{\mathrm{in}}$ and $\mathcal{R}^{\mathrm{out}}$, as shown in FIG. \ref{fig:Ricci_curvature}. The approach is to find the last intersection point between a characteristic starting from radius $r_0$ and the shock surface. Then, we take two nearby characteristics at $r_0\pm \delta r$ (for example, $\delta r = 10^{-5}$) and use the central difference method to compute the required derivatives in (\ref{eq:Ricci curvarute}). The results in FIG. \ref{fig:Ricci_curvature} confirm that the Ricci curvature is discontinuous but finite across the shock surface. As time goes on, $\mathcal{R}^{\mathrm{in}}$ asymptotically approaches zero, consistent with the inner region approaching flat spacetime.

To further study the curvature behavior, we also compute the Kretschmann scalar. In PG coordinates, it is given by:
\begin{equation}\small{
\begin{aligned}
\mathcal{K} &= 4 \left[\left(\partial_{rr} N^x\right)^2 \left(N^x\right)^2+\frac{\left(N^x\right)^4}{r^4}+\right.\\
&\left(2 \partial_{rr} N^x \left(\partial_{r} N^x\right)^2-\frac{4 \partial_t N^x \partial_r N^x}{r^2}-2 \left(\partial_{rr} N^x\right) \partial_{tr}N^x\right) N^x\\
&\left.+\frac{2\left(\partial_t N^x\right)^2+r^2\left(\left(\partial_r N^x\right)^2-\partial_{tr} N^x\right)^2+4 \left(\partial_r N^x\right)^2\left(N^x\right)^2}{r^2}\right].  \nonumber   
\end{aligned} } 
\end{equation}
FIG. \ref{fig:Kretschmann_curvature} shows how the Kretschmann scalar evolves inside and outside the shock surface. The plot illustrates how the curvature changes across the shock surface and highlights the discontinuities and finiteness of the spacetime geometry near the shock surface.

\section{Conclusion and Discussion}\label{conclusion}
In this work, we numerically study the formation and evolution of shock waves in spherically symmetric gravitational collapse within the framework of LQG, using a generalized PG coordinate system. By applying junction conditions, we derive a first-order PDE that determines the location of the junction surface, which also serves as the jump condition for the weak solution. This approach avoids the need for explicit metric solutions on both sides of the junction surface, as the shock surface is solved directly from the PDE without assuming specific interior or exterior geometries. Moreover, our numerical method allows us to handle a PDE with a complex source term and a nontrivial square root structure, capturing key features of the system. The framework developed here can be applied to a wide range of black hole models, offering a practical tool for studying gravitational collapse and exploring quantum gravity effects in highly curved regimes.

Our results ensure a continuous induced metric function across the shock surface, in contrast to earlier findings in \cite{Kelly:2020lec, Husain:2021ojz}. This analysis reveals a coherent spacetime structure that enables us to determine the causal signature of the shock surface. Specifically, we find that the shock surface corresponds to a physical (timelike) shock wave only when the mass is extremely small—on the order of the Planck scale. In these cases, the shock wave remains entirely within the untrapped region, lying inside the inner horizon. For larger black holes, however, the shock surface develops spacelike segments. This highlights fundamental limitations of current effective dynamics in describing such regimes. In particular, quantum tunneling effects may become dominant and necessary to resolve these scenarios. For example, the mass may act as a critical parameter controlling tunneling probabilities between distinct phases of spacetime geometry. The observed structure also provides the possibility of a quantum phase transition across different mass scales, suggesting that higher-order corrections or even fully non-perturbative quantum gravity may be required to describe the near bounce dynamics of macroscopic black holes. These results further highlight the importance of incorporating the evaporation of macroscopic black holes down to Planckian size within existing effective field-theoretic frameworks, in order to consistently describe a coherent and complete spacetime structure across all scales.

Regardless of mass scale, when a black hole forms, the shock wave asymptotically approaches the minimal radius (bounce radius) of the vacuum black hole region. The star's mass becomes increasingly concentrates within the shock, leading the interior region asymptotically flat. In the late-time limit, the shock remains inside the untrapped region, interior to the inner horizon, and does not affect the locations of the inner or outer horizons. This implies that the shock can not be observed by an observer outside the black hole. This leads to a different spacetime structure compare to \cite{Kelly:2020lec, Husain:2021ojz}. The extrinsic curvature and associated curvature invariants across the shock surface remain discontinuous but finite. This is due to the  presence of non-trivial energy-momentum density along the shock and the modification to the standard thin-shell junction conditions for second fundamental forms introduced by the effective dynamics, as discussed in \cite{Giesel:2023hys}. 

Future research directions include extending this analysis to more general gravitational collapse scenarios beyond marginally bound systems and even beyond dust systems, as well as applying the framework to alternative setups such as the asymmetric bounce model proposed in \cite{Giesel:2024mps}. Since physical shock waves appear to be valid only at near-Planckian masses, incorporating black hole evaporation processes is crucial for understanding the evolution of larger black holes to ensure a consistent spacetime structure within effective field theory frameworks. It would also be interesting to explore the quantum behavior beyond the effective framework in regions where the shock surface becomes spacelike in large-mass systems, especially in relation to potential quantum tunneling effects.

\begin{acknowledgments}
The authors thank Kristina Giesel, Carlo Rovelli, Francessca Vidotto and Cong Zhang for useful discussions. H.L. is supported by the start-up funding from Westlake University. H.L. acknowledges the hospitality of Perimeter Institute during his visit. D.Q. acknowledges the hospitality of Westlake University, FAU Erlangen-Nürnberg, and the University of Western Ontario during her visit. Research at Perimeter Institute is supported in part by the Government of Canada through the Department of Innovation, Science and Economic Development Canada and by the Province of Ontario through the Ministry of Colleges and Universities.
\end{acknowledgments}

\bibliographystyle{jhep}
\bibliography{references}

\end{document}